\documentstyle[pra,aps,psfig,preprint]{revtex}

\def\gtsim{\gtrsim}

\begin{document}
\draft

\title{Spontaneous decay of an excited atom in an absorbing dielectric}
\author{S.~Scheel, L. Kn\"oll, and D.-G.~Welsch}
\address{Theoretisch-Physikalisches Institut, 
Friedrich-Schiller-Universit\"at Jena, 
Max-Wien-Platz 1, 07743 Jena, Germany}

\date{\today}
\maketitle

\begin{abstract}
Starting from the quantized version of Maxwell's equations
for the electromagnetic field in an arbitrary linear Kramers-Kronig 
dielectric, spontaneous decay of the excited state of
a two-level atom embedded in a dispersive and absorbing medium
is studied and the decay rate is calculated.
The calculations are performed for both the
(Clausius-Mosotti) virtual cavity model and
the (Glauber-Lewenstein) real cavity model.
It is shown that owing to nonradiative decay associated
with absorption the rate of spontaneous decay
sensitively depends on the cavity radius when the atomic 
transition frequency approaches an absorption band of the  
medium. Only when the effect of absorption is fully 
disregarded, then the familiar local-field correction 
factors are recovered. 
\end{abstract}

\pacs{42.50.-p,42.50.Ct,42.50.Lc}

\section{Introduction}
\label{introduction}

Spontaneous emission is a prime example of the action of vacuum
fluctuations on physically measurable processes. Since the early work of 
Einstein \cite{Einstein} spontaneous emission has been a major
ingredient in the understanding of the effects of what one calls the
vacuum in quantum field theory. The radiation properties of an excited
atom located in free space have been a subject of many studies (for a
comprehensive list of original articles, see, e.g.,
\cite{Milonni94}). 
In particular, the rate of spontaneous emission in free space
(half the Einstein coefficient) is given by
\begin{equation} 
\label{1.1}
\Gamma_{\rm SE} = \Gamma_0 
\equiv \frac{\omega_A^3 \mu^2}{3\pi \hbar \epsilon_0 c^3}\,, 
\end{equation}
where $\omega_A$ is the transition frequency of the atom and $\mu$ is 
the dipole matrix element of the transition. The question 
has been arisen of how a surrounding medium modifies that decay. 
Simple arguments based on the change of the mode density suggest
that the spontaneous emission rate inside a non-absorbing
dielectric should be modified according to \cite{Nienhuis76}
\begin{equation}
\label{1.2}
\Gamma_{\rm SE} = n \Gamma_0,
\end{equation}
where $n$ is the real refractive index of the medium. 
In Eq.~(\ref{1.2}) it is assumed that the local field the
atom interacts with is identical with the 
electromagnetic field in the continuous medium. 
Since in reality the atom is in a small region of free space,
the local field felt by the atom is different from the field
in the continuous medium
\cite{Keller97}, and the decay rate may be expected to be modified to
\begin{equation}
\label{1.3}
\Gamma_{\rm SE} = n \xi \Gamma_0,
\end{equation}
where $\xi$ is the local-field correction factor.
Different models have been used to calculate it.
In the (Clausius-Mosotti) virtual cavity model 
it is given by \cite{Knoester89}
\begin{equation}
\label{1.4}
\xi_{\rm CM} = \left(\frac{n^2+2}{3}\right)^2,
\end{equation}
whereas the (Glauber-Lewenstein) real cavity model
leads to \cite{Glauber91}
\begin{equation}
\label{1.5}
\xi_{\rm GL} = \left(\frac{3n^2}{2n^2+1}\right)^2.
\end{equation}
Recently, experiments have been reported
\cite{Rikken95,Schuurmans98} from which
the real-cavity model may be favored.

As already mentioned, in Eqs.~(\ref{1.2}) -- (\ref{1.5}) it is 
assumed that the refractive index of the medium, which may  
vary with frequency [i.e., $n$ $\!\to$ $\!n(\omega_A)$
in Eqs.~(1.2) -- (1.5)], is real.
However, in reality the refractive index must be a complex
function of frequency, 
\begin{equation}
\label{1.6}
n(\omega) = \eta(\omega) + i\,\kappa(\omega).
\end{equation}
It is well known that causality requires 
the permittivity of the medium, $\epsilon(\omega)$ $\!=$ 
$\!n^2(\omega)$, to be a complex function of frequency
whose real part (responsible for dispersion) and
imaginary part (responsible for absorption) are
related to each other by the Kramers-Kronig relation.
Only when the atomic transition frequency $\omega_A$ is sufficiently 
far from a medium resonance, so that absorption may be 
disregarded, the imaginary part of the refractive index
(at the atomic transition frequency) may
be neglected: $\!n(\omega_A)$ $\!\approx$ $\!\eta(\omega_A)$.

Describing the (undisturbed, continuous) medium in terms of a 
complex permittivity,
in \cite{Barnett92,Barnett96} it is argued that
Eqs.~(\ref{1.3}) -- (\ref{1.5}) can be extended to the spontaneous 
emission of an atom embedded in a lossy dielectric as
\begin{equation}
\label{1.7}
\Gamma_{\rm SE} = \eta(\omega_A) \xi(\omega_A) \Gamma_0 ,
\end{equation}
where the local-field correction factors 
(\ref{1.4}) and (\ref{1.5}) are now regarded as being 
squares of absolute values,   
\begin{equation}
\label{1.8}
\xi_{\rm CM}(\omega_A) = \left|\frac{n^2(\omega_A)+2}{3}\right|^2,
\end{equation}
\begin{equation}
\label{1.9}
\xi_{\rm GL}(\omega_A) 
= \left|\frac{3n^2(\omega_A)}{2n^2(\omega_A)+1}\right|^2 .
\end{equation}
Further, in \cite{Barnett96} the total
decay rate is decomposed as
\begin{equation}
\label{1.10}
\Gamma = \Gamma^\perp + \Gamma^\|,
\end{equation}
where the rates $\Gamma^\perp$ and $\Gamma^\|$, respectively,
are related to the transverse and longitudinal electromagnetic
fields in the medium. The rate $\Gamma^\perp$ is identified with 
the cavity-radius-independent rate
$\Gamma_{\rm SE}$ given by Eq.~(\ref{1.7}), 
and it is argued that the rate $\Gamma^\|$,
which depends on the cavity radius $R$ as $\Gamma^\|$ $\!\sim$ $\!R^{-3}$,
is responsible for nonradiative decay via energy transfer
between the atom and the surrounding (absorbing) dielectric.

>From the study of resonant energy transfer between two guest molecules
in a perfect lattice of absorbing molecules \cite{Juzeliunas94}, 
in \cite{Juzeliunas97} it is argued that 
(within the approximations made) the rate of 
spontaneous emission is given 
by Eq.~(\ref{1.7}) together with Eq.~(\ref{1.8}), i.e., with the 
local-field correction factor that corresponds to the virtual-cavity model. 
However, the total decay rate is purely transverse; i.e., it results only 
from the transverse part of the electromagnetic field in the medium, 
\begin{equation}
\label{1.11}
\Gamma = \Gamma^\perp = \Gamma^{(1)} + \Gamma^{(2)} . 
\end{equation}
It consists of an $R$-independent far-field term $\Gamma^{(1)}$,
which has the form of Eq.~(\ref{1.7}) [together
with Eq.~(\ref{1.8})] and is interpreted as the spontaneous emission 
rate, and a $R$-dependent term $\Gamma^{(2)}$, which
in the near-field zone is proportional to $R^{-3}$ and describes 
nonradiative energy transfer.
    
Recently it has been shown \cite{Scheel98b} that the decay rates 
suggested in \cite{Barnett92,Barnett96} for the virtual-cavity model are 
wrong in general, because the quantum vacuum in the presence of a 
dispersive and absorbing dielectric is not introduced correctly.
The fluctuating part of the polarization field is not fully included 
in the local field coupled to the atom and therefore effects such as
nonradiative energy transfer from the guest atom to the medium via virtual
photon exchange (i.e., transverse-field-assisted energy transfer) 
are omitted. It is just the contribution to the local field of the
fluctuating part of the polarization which gives rise to 
the relevant terms $\sim$ $\!R^{-3}$ and $\sim$ $\!R^{-1}$ in the transverse 
decay rate of an excited atom surrounded by an absorbing medium 
\cite{Scheel98b}. It is worth noting that the results have been 
confirmed within a microscopic approach to the problem  
more recently \cite{fleisch}.

In the virtual-cavity model, the electromagnetic field inside
the cavity, i.e., the local field, is modified by the presence 
of the cavity, but the modification of the field outside the cavity 
is disregarded. Hence the local field introduced in this way 
is not exactly the field that couples to the atom in reality.  
On the contrary, in the real-cavity model the 
mutual modification of the fields outside and inside the 
cavity are taken into account in a consistent way; i.e., 
the atom interacts with a field that exactly satisfies 
both Maxwell's equations and the fundamental commutation rules of 
quantum electrodynamics. It may be therefore expected that
the real cavity model is more suited for describing the
spontaneous decay than the virtual cavity model. 
In particular, the Power-Zienau-Woolley transformation
(see, e.g., \cite{Craig84}) suggests
that (in dipole approximation) only the transverse 
electromagnetic field contributes to the decay rate via  
radiative decay {\em and} nonradiative decay
associated with virtual photon exchange,
the latter being typical for an absorbing medium.

In this article we consider, within the frame of rigorous
quantization of the electromagnetic field in an arbitrary
linear Kramers-Kronig consistent dielectric
\cite{Gruner96,Dung98,Scheel98a}, the spontaneous decay of
an excited atom embedded in an absorbing dielectric, applying
the real-cavity concept.
We find that the rate formulas suggested in \cite{Barnett96}
for the real-cavity model are essentially wrong. At first, only the 
transverse electromagnetic field
contributes to the decay rate, i.e., $\Gamma^\|$ $\!\equiv$ $\!0$,
which contradicts \cite{Barnett96}.
At second, the (purely transverse)
rate not only contains an $R$-independent term but also
terms proportional to $R^{-1}$ and $R^{-3}$ 
which are closely related to nonradiative
decay -- a result which also contradicts \cite{Barnett96}.
As expected, nonradiative decay is only observed
for an absorbing medium. 
It is worth noting that when the atomic transition frequency
is sufficiently far from an absorption band of the medium,
so that absorption may be neglected, our result exactly agrees
with that derived in \cite{Glauber91} for a non-absorbing medium.

The paper is organized as follows. After introducing the
quantization scheme, in Sect.~\ref{scheme} the problem of spontaneous
decay of an exited atom in an absorbing medium is considered.
In Sect.~\ref{CM} the results for decay rate with the virtual cavity model
are outlined, and Sect.~\ref{GL} presents a detailed analysis
of the decay rate with the real cavity model. The results are discussed in
Sect.~\ref{discussion}. Lengthy calculations are given in the Appendix.


\section{Basic equations}
\label{scheme}

Our analysis of the spontaneous decay of an excited atom
embedded in an absorbing medium is based 
on the scheme for quantization of the
electromagnetic field in linear Kramers-Kronig dielectrics
developed in \cite{Gruner96,Dung98,Scheel98a}. 
We start with the phenomenological Maxwell's equations in 
the (temporal) Fourier space, without external sources,
\begin{equation} 
\label{2.1}
{\bf \nabla} \cdot \underline{\hat{\bf B}}({\bf r},\omega) = 0,
\end{equation}
\begin{equation} 
\label{2.1A}
{\bf \nabla} \cdot \left[ \epsilon_0 \epsilon({\bf r},\omega)
\underline{\hat{\bf E}}({\bf r},\omega)\right] =
\underline{\hat{\rho}}({\bf r},\omega) ,
\end{equation}
\begin{equation}
\label{2.1B}
{\bf \nabla} \times \underline{\hat{\bf E}}({\bf r},\omega) = i\omega
\underline{\hat{\bf B}}({\bf r},\omega) , 
\end{equation}
\begin{equation}
\label{2.1C}
{\bf \nabla} \times \underline{\hat{\bf B}}({\bf r},\omega) =
-i\frac{\omega}{c^2} \epsilon({\bf r},\omega) \underline{\hat{\bf
E}}({\bf r},\omega) +\mu_0 \underline{\hat{\bf j}}({\bf r},\omega),
\end{equation}
where $\epsilon({\bf r},\omega)$ $\!=$ $\!\epsilon_R({\bf r},\omega)$
$\!+$ $\!i\epsilon_I({\bf r},\omega)$ is the (spatially varying) 
permittivity satisfying the
Kramers-Kronig relations. When there are no external charges
and currents, then $\underline{\hat{\rho}}({\bf r},\omega)$
and $\underline{\hat{\bf j}}({\bf r},\omega)$, respectively, are the
operator noise charge and current densities that are associated
with absorption according to the dissipation-fluctuation
theorem. They satisfy the equation of continuity,
\begin{eqnarray} 
\label{2.1a}
{\bf \nabla} \cdot \underline{\hat{\bf j}}({\bf r},\omega)
= i \omega \underline{\hat{\rho}}({\bf r},\omega),
\end{eqnarray}
and they are related to the noise polarization
$\underline{\hat{\bf P}}^N({\bf r},\omega)$ as
\begin{eqnarray} 
\label{2.2}
\underline{\hat{\bf j}}({\bf r},\omega) &=& -i\omega
\underline{\hat{\bf P}}^N({\bf r},\omega), 
\\
\label{2.2a}
\underline{\hat{\rho}}({\bf r},\omega) &=& -{\bf \nabla} \cdot
\underline{\hat{\bf P}}^N({\bf r},\omega) . 
\end{eqnarray}
Let $\hat{\bf f}({\bf r},\omega)$ be an infinite set
of bosonic field operators which may be viewed as being collective 
excitations of the electromagnetic field, the medium polarization, 
and the reservoir. All operators in the theory can then be 
expressed in terms of these basic field operators using
the relation
\begin{equation} 
\label{2.3}
\underline{\hat{\bf j}}({\bf r},\omega) = \omega \sqrt{\frac{\hbar
\epsilon_0}{\pi} \epsilon_I({\bf r},\omega)} 
\,\hat{\bf f}({\bf r},\omega) .
\end{equation}

In particular, from Maxwell's equations the electric field
(in Fourier space)
is given by a convolution with the classical dyadic Green function,
\begin{equation} 
\label{2.5}
\underline{\hat{E}}_k({\bf r},\omega) = i\mu_0 \int d^3{\bf r}'
\,\omega G_{kk'}({\bf r},{\bf r}',\omega) 
\underline{\hat{j}}_{k'}({\bf r}',\omega) ,
\end{equation}
where $G_{kk'}({\bf r},{\bf r}',\omega)$ satisfies the partial
differential equation 
\begin{equation} 
\label{2.4}
\left[ \partial_i^r \partial_k^r - \delta_{ik} \left( \triangle^r
+ \frac{\omega^2}{c^2} \epsilon({\bf r},\omega) \right) \right]
G_{kk'}({\bf r},{\bf r'},\omega)
= \delta_{ik'} \delta({\bf r}-{\bf r'}) .
\end{equation}
Integration with respect to $\omega$
then yields the operator of the electric field as
\begin{equation}
\label{2.6}
\hat{\bf E}({\bf r}) = \hat{\bf E}^{(+)}({\bf r})
+ \hat{\bf E}^{(-)}({\bf r}),
\quad
\hat{\bf E}^{(-)}({\bf r})
= \left[\hat{\bf E}^{(+)}({\bf r})\right]^\dagger,
\end{equation}
\begin{equation} 
\label{2.7}
\hat{E}_k^{(+)}({\bf r}) = 
\int_{0}^{\infty} d\omega\,\underline{\hat{E}}_k({\bf r},\omega) 
= i\mu_0 \int_{0}^{\infty} d\omega \int d^3{\bf r}'
\,\omega G_{kk'}({\bf r},{\bf r}',\omega) 
\underline{\,\hat{\!j}}_{k'}({\bf r}',\omega) .
\end{equation}
Substituting in Eq.~(\ref{2.7}) for the
current density the expression given in Eq.~(\ref{2.3}) yields
the electric field in terms of the bosonic basic 
fields. It can be proven \cite{Scheel98a} that the quantization scheme
is fully consistent with QED for arbitrary linear dielectrics, i.e.,
\begin{equation} 
\label{2.7a}
\epsilon_0 \left[ \hat{E}_k({\bf r}) , \hat{B}_l({\bf r'}) \right] 
= -i\hbar \epsilon_{klm} \partial_m^r \delta({\bf r}-{\bf r'}) ,
\end{equation}
\begin{equation} 
\label{2.7b}
\left[ \hat{E}_k({\bf r}), \hat{E}_l({\bf r'})\right] 
= \left[ \hat{B}_k({\bf r}), \hat{B}_l({\bf r'})
\right] = 0.
\end{equation}

The electric and magnetic fields can be of course expressed in terms
of vector ($\hat{\bf A}$) and scalar ($\hat{\varphi}$)
potentials. In what follows we will set the scalar potential
equal to zero. This gauge condition implies that both the transverse 
and the longitudinal electric fields are obtained from the 
vector potential
\begin{equation} 
\label{2.8}
\hat{\bf A}({\bf r}) = \hat{\bf A}^{(+)}({\bf r})
+ \hat{\bf A}^{(-)}({\bf r}),
\end{equation}
\begin{equation} 
\label{2.9}
\hat{A}_k^{(+)}({\bf r})   
= \mu_0 \int_{0}^{\infty} d\omega \int d^3{\bf r}'
\, G_{kk'}({\bf r},{\bf r}',\omega) 
\underline{\,\hat{\!j}}_{k'}({\bf r}',\omega) .
\end{equation} 

Let us now consider the case when an external (two-level) atomic
system at position ${\bf r}_A$ is present. 
Treating the interaction of such a guest atom with the 
electromagnetic field in dipole and rotating wave approximations, the
Hamiltonian  of the total system can be given by
\begin{equation}
\label{2.10}
\hat{H} = \int d^3{\bf r} \int_0^\infty d\omega \,\hbar\omega
\hat{\bf f}^\dagger({\bf r},\omega) \hat{\bf f}({\bf r},\omega)
+\sum_{\alpha=1}^{2}\hbar\omega_\alpha\hat{A}_{\alpha\alpha}
-\left[
i\omega_{21}\hat{A}_{21}\, \hat{\bf A}^{(+)}({\bf r}_A)\!\cdot\!{\bf d}_{21}
+{\rm H.c.}\right].
\end{equation}
Here the atomic operators $\hat{A}_{\alpha\alpha'}$ 
$\!=$ $\!|\alpha\rangle\langle \alpha'|$
are introduced, with $|\alpha\rangle$ being the energy eigenstates 
of the guest atom
\mbox{($\alpha$ $\!=$ $\!1,2$)}. The energies of the two states are
$\hbar\omega_1$ and $\hbar\omega_2$ ($\hbar\omega_2$ $\!>$
$\!\hbar\omega_1$), and $\omega_{21}$ $\!=$
$\omega_2$ $\!-$ $\!\omega_1$ and ${\bf d}_{21}$, respectively,
are the atomic transition frequency and dipole moment.
Note that in the interaction term in Eq.~(\ref{2.10}) the 
$\hat{\bf A}^2$ term and the counter-rotating terms have
been dropped. 

In the Heisenberg picture the
equations of motion then read as, on recalling Eqs.~(\ref{2.3})
and (\ref{2.9}),
\begin{eqnarray}
\label{2.11}
\,\dot{\hat{\!A}}_{22} &=& 
- \frac{\omega_{21}}{\hbar} \hat{A}_{21} \,
\hat{\bf A}^{(+)}({\bf r}_A)\!\cdot\!{\bf d}_{21}\,  + {\rm H.c.},
\\
\label{2.11a}
\,\dot{\hat{\!A}}_{11} &=& -\,\dot{\hat{\!A}}_{22}\,,
\end{eqnarray}
\begin{equation}
\label{2.12}
\,\dot{\hat{\!A}}_{21} = i \omega_{21} \hat{A}_{21}
+ \frac{\omega_{21}}{\hbar} \hat{\bf A}^{(-)}({\bf r}_A)\!\cdot\!{\bf d}_{21}\,
\left(\hat{A}_{22}-\hat{A}_{11}\right), 
\end{equation}
\begin{equation}
\label{2.13}
\,\dot{\hat{\!f}}_i({\bf r},\omega) = - i\omega \hat{f}_i({\bf r},\omega)
+\, \frac{\omega_{21}\omega}{c^2} 
\sqrt{\frac{\epsilon_I({\bf r},\omega)}{\hbar\pi\epsilon_0}}
\, (d_{21})_k G_{ki}^\ast({\bf r}_A,{\bf r},\omega)\,\hat{A}_{12}\,.
\end{equation}
Substituting in the vector potential in Eqs.~(\ref{2.11}) -- 
(\ref{2.12}) for $\hat{f}_i({\bf r},\omega,t)$ the formal
solution of Eq.~(\ref{2.13}), i.e.,
\begin{equation}
\label{2.13a}
\hat{A}^{(+)}_i({\bf r},t) = \hat{A}^{(+)}_{{\rm free}\,i}({\bf r},t)
+\, \frac{\omega_{21}}{\pi \epsilon_0 c^2}  (d_{21})_k \int_0^\infty
\!d\omega \, 
\bigg[
{\rm Im}\,G_{ik}({\bf r},{\bf r}_A,\omega)
\int_{t'}^t \!d\tau\,
{\rm e}^{-i\omega (t-\tau)} \hat{A}_{12}(\tau) 
\bigg],
\end{equation}
a system of integro-differential 
equations for the atomic quantities is obtained.
[Note that Eq.~(\ref{B4}) has been used for
deriving Eq.~(\ref{2.13a}).]
At this stage a Markov approximation can be introduced, 
and the integro-differential equations reduce to 
Langevin-type differential equations    
(Appendix \ref{markov})
\begin{eqnarray}
\label{2.15}
\,\dot{\hat{\!A}}_{22}
&=& - \Gamma \hat{A}_{22} 
- \left[\hat{A}_{21}\, \frac{\omega_{21}}{\hbar} 
\hat{\bf A}_{\rm free}^{(+)}({\bf r}_A,t)\!\cdot\!{\bf d}_{21}
+ {\rm H.c.} \right],
\\
\label{2.15a}
\,\dot{\hat{\!A}}_{11} &=& -\,\dot{\hat{\!A}}_{22}\,,
\end{eqnarray}
\begin{equation}
\label{2.16}
\,\dot{\hat{\!A}}_{21} = 
\left[i(\omega_{21} - \delta\omega) - {\textstyle\frac{1}{2}\Gamma} \right]
\hat{A}_{21} 
+ \,\frac{\omega_{21}}{\hbar} 
\hat{\bf A}_{\rm free}^{(-)}({\bf r}_A,t)\!\cdot\!{\bf d}_{21}\,
\left(\hat{A}_{22}-\hat{A}_{11}\right), 
\end{equation}
where $\Gamma$ is the rate of spontaneous decay of the excited state 
of the guest atom,
\begin{eqnarray}
\label{2.17}
\Gamma = \frac{2\omega_A^2\mu_k\mu_{k'}}{\hbar\epsilon_0c^2}\,
{\rm Im}\,G_{kk'}({\bf r}_A,{\bf r}_A,\omega_A)
\end{eqnarray}
[$\mu_k$ $\!\equiv$ $\!(d_{21})_k$, $\omega_A$ $\!\equiv$ $\!\omega_{21}$],
and $\delta\omega$ is the (contribution of 
the dielectric to the) Lamb shift [see Eq.~(\ref{B7})].
Note that $\hat{\bf A}_{{\rm free}}^{(\pm)}({\bf r},t)$ evolves 
freely. From Eq.~(\ref{2.5}) together with Eqs.~(\ref{2.3}) and 
(\ref{B4}) it can be proved that the quantization scheme
exactly yields, in agreement with the dissipation-fluctuation theorem,
the relation \cite{Abrikosov}
\begin{equation} 
\label{2.18}
{\rm Im}\,G_{kk'}({\bf r},{\bf r}',\omega')\;\delta(\omega-\omega') 
= \frac{\pi \epsilon_0 c^2}{\hbar \omega^2}
\,\langle 0 | \Big[\underline{\hat{E}}_k({\bf r},\omega),
\underline{\hat{E}}_{k'}^\dagger({\bf r}',\omega')\Big] | 0 \rangle . 
\end{equation}

As long as the Markov approximation applies,
the spontaneous decay can be described in terms
of the rate (\ref{2.17}), the rate formula being 
valid for arbitrary dielectrics and geometries.
Especially, for an atom in vacuum we have 
\begin{equation} 
\label{2.19}
{\rm Im} \,G_{kk'}({\bf r}_A,{\bf r}_A,\omega_A) =
\frac{\omega_A}{6\pi c} \, \delta_{kk'}
\end{equation}
[see Eqs.~(\ref{C1}) -- (\ref{C5}) for $\epsilon$ $\!=$ $\!1$],
which leads to the well-known result (\ref{1.1}), 
\begin{equation} 
\label{2.20}
\Gamma = \Gamma_0 = \frac{\omega_A^3 \mu^2}{3\pi \hbar \epsilon_0 c^3} \,.
\end{equation}
A guest atom in a dielectric is situated in a
small free-space region and is surrounded by medium atoms.
Frequently a cavity model is used for describing the
situation. An atom in an empty cavity in an otherwise continuous 
medium is considered and it is assumed that the linear
dimensions of the cavity are much less then the atomic transition 
wavelength. In particular, for isotropic systems a spherical cavity 
of radius $R$ may be considered. With regard to Eq.~(\ref{2.17}), the
``only'' problem that remains is the calculation of (the imaginary part 
of) the classical Green tensor for a dielectric medium
of given permittivity which is disturbed by a
small free-space inhomogeneity.


\section{Virtual cavity model}
\label{CM}

In the virtual cavity model it is assumed that the field outside
the sphere is not modified by the small region of free space
inside the sphere, and the (local) electric field 
$\underline{\bf E}'({\bf r},\omega) $ inside the sphere is given by
\cite{bornwolf}
\begin{equation} 
\label{3.1}
\underline{\hat{\bf E}}'({\bf r},\omega) 
= \underline{\hat{\bf E}}({\bf r},\omega) 
+ \frac{1}{3\epsilon_0} \underline{\hat{\bf P}}({\bf r},\omega) ,
\end{equation}
where $\underline{\bf E}({\bf r},\omega)$ and
$\underline{\bf P}({\bf r},\omega)$, respectively, are the 
electric and polarization fields 
in the unperturbed continuous medium. From 
Maxwell's equations (\ref{2.1A}) and (\ref{2.1C}) 
together with Eqs.~(\ref{2.1a}) -- (\ref{2.3}) it is seen that
\begin{equation} 
\label{3.2}
\underline{\hat{\bf P}}({\bf r},\omega) 
= \epsilon_0 \left[\epsilon({\bf r},\omega) -1\right] 
\underline{\hat{\bf E}}({\bf r},\omega) +
\underline{\hat{\bf P}}{^N}({\bf r},\omega) ,
\end{equation}
where 
\begin{equation} 
\label{3.2a}
\underline{\hat{\bf P}}{^N}({\bf r},\omega)
= i \sqrt{\frac{\hbar\epsilon_0}{\pi}\,\epsilon_I({\bf r},\omega)}
\,\hat{\bf f}({\bf r},\omega)
\end{equation}
is the noise polarization associated with absorption. For
classical optical fields at room temperatures the noise 
polarization weakly contributes to the polarization and the 
local field, and therefore it may be neglected. Obviously, for 
quantum fields and especially for the quantum vacuum, whose 
coupling to the guest atom gives rise to the spontaneous decay, the noise 
polarization must not be omitted, because it is nothing other but a
part of the quantum vacuum. Combining Eqs.~(\ref{3.1}) and (\ref{3.2}) 
yields the local-field operator
\begin{equation} 
\label{3.3}
\underline{\hat{\bf E}}'({\bf r},\omega)
= \frac{1}{3}
[\epsilon({\bf r},\omega) +2] \underline{\hat{\bf E}}({\bf r},\omega)
+\frac{1}{3\epsilon_0} \underline{\hat{\bf P}}{^N}({\bf r},\omega) .
\end{equation}
It can be shown that the local electromagnetic field satisfies 
the equal-time commutation relations \cite{Scheel98b}
\begin{equation} 
\label{3.4}
\epsilon_0 \left[ \hat{E}_k'({\bf r}) , \hat{B}_l'({\bf r'}) \right] 
= -i\hbar \epsilon_{klm} \partial_m^r \delta({\bf r}-{\bf r'}) 
\left\{1+\textstyle{\frac{1}{9}} [\epsilon({\bf r},0) -1] \right\},
\end{equation}
\begin{equation} 
\left[ \hat{E}_k'({\bf r}), \hat{E}_l'({\bf r'})\right] 
= \left[ \hat{B}_k'({\bf r}), \hat{B}_l'({\bf r'})
\right] = 0,
\end{equation}
Comparing with the correct commutation relations, 
we see that the virtual cavity model may be 
regarded as being consistent with QED (over the whole
frequency domain), provided that 
\begin{equation} \label{3.5}
\epsilon({\bf r},0) \ll 10;
\end{equation}
i.e., the value of the static permittivity must not be too large. 

Now we can turn to the calculation of the spontaneous decay rate,
Eq.~(\ref{2.17}). Recalling Eq.~(\ref{2.18}), we may write
\begin{equation} 
\label{3.6}
{\rm Im}\,G_{kk'}({\bf r},{\bf r}',\omega_A)\;\delta(\omega-\omega_A) 
= \frac{\pi \epsilon_0 c^2}{\hbar \omega^2}
\,\langle 0 | \Big[\underline{\hat{E}}_k'({\bf r},\omega),
\underline{\hat{E}}_{k'}'^\dagger({\bf r}',\omega_A)\Big] | 0 \rangle 
\end{equation}
with $|{\bf r}$ $\!-$ $\!{\bf r}_A|$, $\!|{\bf r}'$ $\!-$ $\!{\bf r}_A|$ 
$\!<$ $\!R$ and $\hat{\bf E}'$ from Eq.~(\ref{3.3}). Since 
$\underline{\hat{\bf E}}$ in Eq.~(\ref{3.3}) is determined 
by Eq.~(\ref{2.5})
with the Green tensor for the 
field in the undisturbed continuous medium, knowledge of 
the imaginary part of 
that Green tensor is sufficient to calculate 
the decay rate.
However, for \mbox{${\bf r},{\bf r}'$ $\!\to$ $\!{\bf r}_A$} 
a singular contribution
to the rate is observed, which reflects the fact that 
the description of the dielectric as a continuous medium
contradicts a precise determination of the position of the 
guest atom. 
The problem might be overcome by regularization, e.g., 
by averaging Eq.~(\ref{3.6})
over the sphere. Combining Eqs.~(\ref{2.17}) and (\ref{3.6})
and using Eq.~(\ref{3.3}) [together with Eqs.~(\ref{2.3}),
(\ref{2.5}), and (\ref{3.2a})] yields \cite{Scheel98b}
\begin{eqnarray} 
\label{3.6A}
\lefteqn{
\Gamma_{\rm CM} = \frac{2\omega_A^2 \mu_k \mu_{k'}}{\hbar \epsilon_0 c^2}
\left| \frac{\epsilon(\omega_A)+2}{3}
\right|^2  {\rm Im}\, 
\overline{G_{kk'}^{\rm M}({\bf r},{\bf r}',\omega)}
}
\nonumber \\&&\hspace{2ex}
+\,\frac{4\omega_A^2}{3\hbar \epsilon_0 c^2} 
\epsilon_I(\omega_A) \mu_k \mu_{k'} \,
{\rm Re}\!\left[
\frac{\epsilon(\omega_A)+2}{3} \;
\overline{G_{kk'}^{\rm M}({\bf r},{\bf r}',\omega_A)} 
\right]
+\,\frac{2}{9\hbar \epsilon_0}  \epsilon_I(\omega_A) \mu_k \mu_{k'}
\overline{\delta_{kk'} \delta({\bf r}-{\bf r}')} 
\end{eqnarray}
(the bar introduces averaging over the sphere), where 
$G_{kk'}^{\rm M}({\bf r},{\bf r}',\omega)$ is the Green tensor 
of the mean field in the undisturbed medium, 
and $\epsilon(\omega_A)$ $\!\equiv$ 
$\!\epsilon({\bf r}_A,\omega_A)$. Note that the permittivity
can be assumed to be constant over the small sphere.
The first term in Eq.~(\ref{3.6A}) corresponds to the
result obtained in \cite{Barnett92,Barnett96}, without taking 
account of the contribution of the noise polarization to the 
quantum vacuum. The noise polarization gives rise to the second 
term and the third term in Eq.~(\ref{3.6A})  -- terms that are
proportional to $\epsilon_I(\omega_A)$ and typically observed for 
absorbing media. 

When the position of the guest atom in the medium is sufficiently 
far from inhomogeneities (such as the surface of the dielectric body)
the Green tensor $G_{kk'}^{\rm M}({\bf r},{\bf r}',\omega)$
in Eq.~(\ref{3.6A}) may be identified with that for bulk material
as given in Appendix \ref{greenbulk}.
Inserting for $G_{kk'}^{\rm M}({\bf r},{\bf r}',\omega_A)$ in 
Eq.~(\ref{3.6A}) the result of 
Eqs.~(\ref{C1}), (\ref{C2}), and (\ref{C5})  
and averaging with respect to ${\bf r}$ and ${\bf r}'$ separately
over a sphere, on assuming equidistribution,
we derive \cite{note} 
\begin{equation}
\label{3.7}
\Gamma_{\rm CM} = \Gamma_{\rm CM}^\| + \Gamma_{\rm CM}^\perp,
\end{equation}
where $\Gamma_{\rm CM}^\|$ and $\Gamma_{\rm CM}^\perp$,
respectively, are related to the longitudinal and transverse 
parts of the Green tensor, 
\begin{equation} 
\label{3.7b}
\Gamma_{\rm CM}^\| =
\Gamma_0 \, \frac{4\epsilon_I(\omega_A)}{27|\epsilon(\omega_A)|^2}
\left( \frac{c}{\omega_A R} \right)^3, 
\end{equation}
\begin{eqnarray} 
\label{3.7a}
\lefteqn{
\Gamma_{\rm CM}^\perp = \Gamma_0 \Bigg\{ \eta(\omega_A) 
\Bigg[ \Bigg|
\frac{\epsilon(\omega_A)+2}{3} \Bigg|^2 -
\frac{2\epsilon_I^2(\omega_A)}{9} \Bigg]  
}
\nonumber \\ &&\hspace{4ex} 
+ \,\epsilon_I(\omega_A) \left[\epsilon_R(\omega_A) +2 \right] \left[
\frac{8}{15} \left( \frac{c}{\omega_A R} \right)- \frac{2}{9}
\kappa(\omega_A) \right] 
+ \,\frac{25\epsilon_I(\omega_A)}{54} \left( \frac{c}{\omega_A R}
\right)^3 \Bigg\} + {\cal O}(R)
\end{eqnarray}
($|R\sqrt{\epsilon(\omega_A)}\,\omega_A/c|$ 
$\!\ll$ $\!1$),
with $\Gamma_0$ being the free-space spontaneous emission rate defined
in Eq.~(\ref{1.1}). From inspection of 
Eqs.~(\ref{3.7}) -- (\ref{3.7a})
it is seen that, when absorption can be disregarded, i.e.,
$\epsilon_I(\omega_A)$ $\!\approx$ $\!0$ and hence
$\epsilon(\omega_A)$ $\!\approx$ $\!\epsilon_R(\omega_A)$,
$n(\omega_A)$ $\!\approx$ $\!\sqrt{\epsilon_R(\omega_A)}$,
then $\Gamma_{\rm CM}$ $\!\approx$ $\!\Gamma_{\rm CM}^\perp$ 
reduces to $\Gamma_{\rm SE}$ given in Eq.~(\ref{1.3}) with
the local-field correction factor (\ref{1.4}). It is further
seen that for absorbing media the rate $\Gamma_{\rm CM}^\perp$ 
becomes quite different from that given in Eq.~(\ref{1.7}) with the 
local-field correction factor (\ref{1.8}), because
of the effect of the noise polarization.
For more details, the reader is referred to \cite{Scheel98b}. 
Most recently, a more microscopic derivation of the decay rate
has yielded, apart from regularization factors, 
the same results \cite{fleisch}.

It is worth noting that the 
$R$-dependent terms in Eq.~(\ref{3.7a}) solely result from the noise 
polarization. In particular, the term $\sim R^{-3}$ may be regarded
as describing nonradiative decay via dipole-dipole energy
transfer from the guest atom to the surrounding medium. From 
Eqs.~(\ref{3.7}) -- ~(\ref{3.7a}) it is seen
that the terms $\!\sim$ $\!R^{-3}$ can be combined to obtain
an overall rate for the nonradiative dipole-dipole energy
transfer. Obviously, the decomposition of $\Gamma_{\rm CM}$ in 
$\Gamma_{\rm CM}^\perp$ and $\Gamma_{\rm CM}^\|$ has nothing
to do with a decomposition in radiative and nonradiative
decay channels in general. 

It should be pointed out that the averages
in Eq.~(\ref{3.6A}), which correspond to 
regularization at ${\bf r}$ $\!\to$ $\!{\bf r}'$,
can be taken in different ways. In other words,
the $R$-dependent terms in Eqs.~(\ref{3.7b})
and (\ref{3.7a}) are determined only up to some regularization 
factors. Hence, not only the the cavity radius $R$ but also
the scaling factors of the absorption-assisted $\sim$ $\!R^{-1}$ and
$\sim$ $\!R^{-3}$ terms are undetermined in the model.


\section{Real cavity model}
\label{GL}

In the real cavity model the exact Green tensor for the system
disturbed by a small free-space inhomogeneity is inserted
in the rate formula (\ref{2.17}). In other words, the
electromagnetic field inside and outside the cavity exactly
solves Maxwell's equations (\ref{2.1}) -- (\ref{2.1C}) together 
with the standard boundary conditions at the surface of the cavity.
In contrast to the virtual cavity approach, in the real cavity
approach the field inside the cavity exactly satisfies the
fundamental QED equal-time commutation relations (\ref{2.7a}) 
and (\ref{2.7b}), and the Green tensor does not lead to a 
singular contribution to the decay rate. 
The Green tensor for an inhomogeneous problem of that type can 
always be written as a sum of the Green tensor for a homogeneous
problem and some tensor that obeys a source-free wave equation
and ensures the boundary conditions to be satisfied \cite{chew}.
Since the guest atom is situated in an empty cavity,
the relevant Green tensor reads as 
\begin{equation}
\label{4.0a}
G_{kk'}({\bf r},{\bf r}_A,\omega_A)
= G_{kk'}^{\rm V}({\bf r},{\bf r}_A,\omega_A)
+ \tilde{G}_{kk'}({\bf r},{\bf r}_A,\omega_A)
\quad ({\bf r} \to {\bf r}_A)
\end{equation}
where $G_{kk'}^{\rm V}({\bf r},{\bf r}_A,\omega_A)$ is simply the
vacuum Green tensor, which is given by 
Eqs.~(\ref{C1}) -- (\ref{C3})
with \mbox{$\epsilon(\omega)$ $\!=$ $\!1$}, and 
$\tilde{G}_{kk'}({\bf r},{\bf r}_A,\omega_A)$ describes the
effect of reflection at the cavity surface. Obviously,
$G_{kk'}^{\rm V}({\bf r},{\bf r}_A,\omega_A)$ has no
longitudinal imaginary part, 
\begin{equation}
\label{4.0b}
{\rm Im}\,G_{kk'}^{{\rm V}\|}({\bf r},{\bf r}_A,\omega_A) = 0
\qquad ({\bf r} \to {\bf r}_A)
\end{equation}
Since the tensor $\tilde{G}_{kk'}({\bf r},{\bf r}_A,\omega_A)$ is 
related to a source-free problem, it is transverse, and hence
\begin{equation}
\label{4.0c}
\tilde{G}_{kk'}^\|({\bf r},{\bf r}_A,\omega_A) = 0
\qquad ({\bf r} \to {\bf r}_A).
\end{equation}
The imaginary part of $G_{kk'}({\bf r}_A,{\bf r}_A,\omega_A)$ 
is therefore equal to the imaginary part of the transverse part of the
Green tensor, so that the rate formula (\ref{2.17}) in the real 
cavity model reads  
\begin{eqnarray}
\label{4.0d}
\Gamma_{\rm GL} = \frac{2\omega_A^2\mu_k\mu_{k'}}{\hbar\epsilon_0c^2}\,
{\rm Im}\,G_{kk'}^\perp({\bf r}_A,{\bf r}_A,\omega_A).
\end{eqnarray}
In other words, in the real cavity model the longitudinal
field does not contribute to the decay rate. Thus, the
longitudinal decay rate $\Gamma_{\rm GL}^\|$ given in 
\cite{Barnett96} is an artifact.

In order to calculate $\Gamma_{\rm GL}$ further, let us again consider
a spherical cavity of radius $R$ in bulk material, with the guest atom
being situated at the center of the sphere. The Green
tensor for a spherical two-layer system is given
in Appendix~\ref{green}. From Eqs.~(\ref{4.0a}) -- (\ref{4.0c})
together with 
Eq.~(\ref{C5}) [for $\epsilon(\omega)$ $\!=$ $\!1$]
and Eq.~(\ref{A5}) it follows that
\begin{equation} 
\label{4.a1}
{\rm Im}\, G_{kk'}^\perp({\bf r}_A,{\bf r}_A,\omega_A) 
= \frac{\omega_A}{6\pi c} \left[ 1+ {\rm Re}\,C_1^N(\omega_A) \right]
\delta_{kk'}, 
\end{equation}
with the reflection coefficient $C_1^N(\omega_A)$ being given by
Eq.~(\ref{A6}). Hence, for a spherical cavity the spontaneous decay 
rate (\ref{4.0d}) takes the form of 
\begin{equation} 
\label{4.a2}
\Gamma_{\rm GL} 
= \Gamma_0 \left[ 1+ {\rm Re}\,C_1^N(\omega_A) \right], 
\end{equation}
where $\Gamma_0$ is the free-space spontaneous emission 
rate (\ref{1.1}). The reflection coefficient $C_1^N(\omega_A)$
in Eq.~(\ref{4.a2}) is a function of $R$
and given in Eq.~(\ref{A6}) explicitly. For $\omega_A R/c$ 
$\!=$ $\!2\pi R/\lambda_A$ $\!\ll$ $\!1$ 
we expand it in powers of $R$ to obtain
\begin{eqnarray}
\label{4.a5}
C_1^N(\omega_A) 
&=& - \frac{3i[\epsilon(\omega_A)\!-\!1]}{2\epsilon(\omega_A)\!+\!1} 
\left(\frac{c}{\omega_A R}\right)^3
-\, \frac{9i[4\epsilon^2(\omega_A)\!-\!3\epsilon(\omega_A)\!-\!1]}
{5[2\epsilon(\omega_A)+1]^2} 
\left( \frac{c}{\omega_A R} \right)
+ \, \frac{9\epsilon^{5/2}(\omega_A)}{[2\epsilon(\omega_A)+1]^2}
\nonumber \\ && \hspace{5ex}
-1 +{\cal O}(R),
\end{eqnarray}
from which it follows that
\begin{eqnarray} 
\label{4.a6}
\lefteqn{
\Gamma_{\rm GL}
= \Gamma_0\Bigg\{
\frac{9\epsilon_I(\omega_A)}{|2\epsilon(\omega_A)\!+\!1|^2}
\left( \frac{c}{\omega_A R} \right)^3
+\,\frac{9\epsilon_I(\omega_A)
[28|\epsilon(\omega_A)|^2\!+\!12\epsilon_R(\omega_A)\!+\!1]}
{5|2\epsilon(\omega_A)+1|^4 } 
\left( \frac{c}{\omega_A R} \right)
}
\nonumber \\ &&\hspace{2ex}
+\,\frac{9\eta(\omega_A)}{|2\epsilon(\omega_A)+1|^4} 
\Big[
4|\epsilon(\omega_A)|^4 +4\epsilon_R(\omega_A) |\epsilon(\omega_A)|^2
+\,\epsilon_R^2(\omega_A) -\epsilon_I^2(\omega_A) 
\Big] 
\nonumber \\ &&\hspace{2ex}
-\,\frac{9\kappa(\omega_A)\epsilon_I(\omega_A)}{|2\epsilon(\omega_A)+1|^4}
\left[ 4 |\epsilon(\omega_A)|^2 +2
\epsilon_R(\omega_A) \right] 
\Bigg\} 
+ {\cal O}(R) .
\end{eqnarray}

Needless to say that when setting $\epsilon(\omega)$ $\!=$ $\!1$,
then the free-space spontaneous emission rate is recovered.
When the atomic transition frequency is far from
an absorption band of the medium, then absorption may be disregarded, 
i.e., $\epsilon_I(\omega_A)$ $\!\approx$ $\!0$ [and hence 
$\epsilon(\omega_A)$ $\!\approx$ $\!\epsilon_R(\omega_A)$,
$n(\omega_A)$ $\!\approx$ $\!\sqrt{\epsilon_R(\omega_A)}$]. From
inspection of Eq.~(\ref{4.a6}) we see that for 
$\epsilon_I(\omega_A)$ $\!\to$ $\!0$ the 
term proportional to $R^0$ is the leading term, which exactly
gives rise to the rate formula (\ref{1.3}) together with the 
correction factor (\ref{1.5}), i.e., we recover the familiar 
result derived in \cite{Glauber91} for real refractive index. 
We further see that for an absorbing medium the rate formula
cannot be given in the form of Eq.~(\ref{1.7}) together with
Eq.~(\ref{1.8}), as is suggested in \cite{Barnett96}.
Equation (\ref{4.a6}) reveals that for an absorbing medium
terms proportional to $R^{-3}$ and $R^{-1}$ are observed,
so that the decay rate sensitively depends on the radius of the sphere.
In particular, the near-field term proportional to $R^{-3}$ can 
again be regarded as corresponding to nonradiative decay via dipole-dipole
energy transfer from the guest atom to the medium.  

It should be pointed out that the
condition that \mbox{$\omega_AR/c$ $\!\ll$ $\!1$}; i.e., 
the (optical) wavelength $\lambda_A$ of the atomic transition 
must be large compared with
the radius $R$ of the cavity, is in full agreement with
the Markov approximation used in order to introduce
a decay rate. From inspection of Eq.~(\ref{A6}) it is seen 
that the (real part of the) reflection coefficient 
$C_1^N(\omega)$ becomes a rapidly varying function 
of frequency for $\omega R/c$ $\!\gtsim$ $\!1$, and hence the
Markov approximation fails. In that case the
sphere acts like a micro-cavity resonator and
memory effects must be included in the temporal
evolution of the atom, which prevents the
excited state from decaying exponentially.


\section{Discussion}
\label{discussion}

\begin{figure}[h]
\hspace{2cm}
\psfig{file=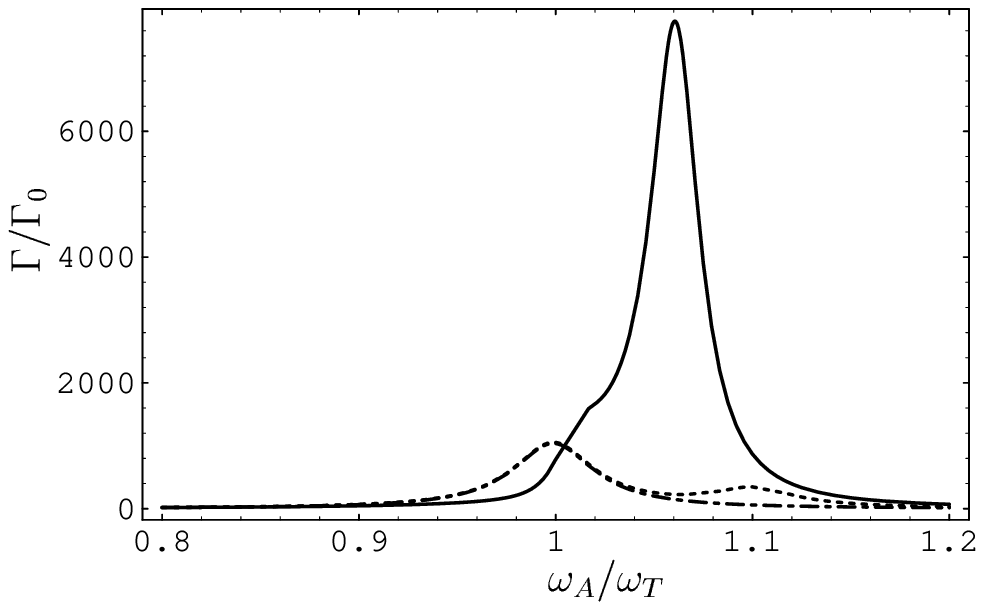,width=10cm}
\caption{\label{fig1} 
The spontaneous decay rate $\Gamma$, Eq.~(\protect\ref{2.17}),
is shown as a function of the atomic transition frequency $\omega_A$
near a medium resonance for the model permittivity (\protect\ref{4.1}) 
(\mbox{$\omega_P$ $\!=$ $\!0.46\,\omega_T$}, $\gamma$ $\!=$
$\!0.05\,\omega_T$) and $R$ $\!=$ $\!0.02\,\lambda_A$.
The solid line corresponds to the real-cavity model, $\Gamma_{\rm GL}$
from Eq.~(\protect\ref{4.a6}), and the dotted line corresponds to
the virtual-cavity model, $\Gamma_{\rm CM}$
from Eq.~(\protect\ref{3.7}), the broken line indicating
the transverse-field assisted rate $\Gamma_{\rm CM}^\perp$
from Eq.~(\protect\ref{3.7a}).
}
\end{figure}

\begin{figure}[h]
\hspace{2cm}
\psfig{file=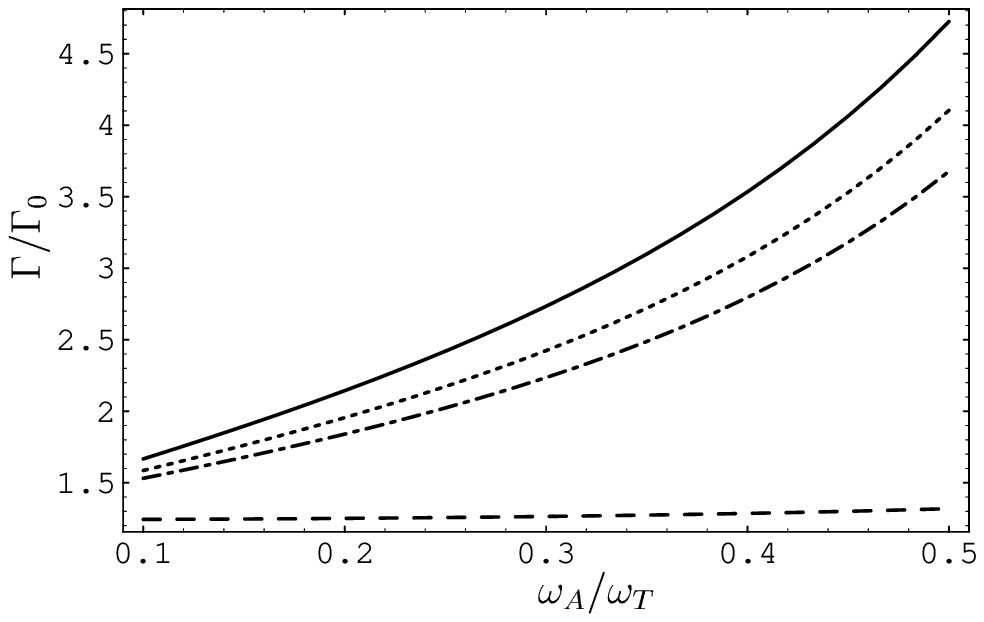,width=10cm}
\caption{\label{fig2}
The spontaneous decay rate $\Gamma$, Eq.~(\protect\ref{2.17}),
is shown as a function of the atomic transition frequency $\omega_A$
far from a medium resonance for the model permittivity (\protect\ref{4.1}) 
(\mbox{$\omega_P$ $\!=$ $\!0.46\,\omega_T$}, $\gamma$ $\!=$
$\!0.05\,\omega_T$) and $R$ $\!=$ $\!0.02\,\lambda_A$.
The solid line corresponds to the real-cavity model, $\Gamma_{\rm GL}$
from Eq.~(\protect\ref{4.a6}), and the dotted line corresponds to
the virtual-cavity model, $\Gamma_{\rm CM}$
from Eq.~(\protect\ref{3.7}), the broken line indicating
the transverse-field assisted rate $\Gamma_{\rm CM}^\perp$
from Eq.~(\protect\ref{3.7a}). For comparison,
the rate $\Gamma_{\rm GL}$ as obtained from 
Eq.~(\protect\ref{1.3}) together with (\protect\ref{1.5}) is
shown (dashed line).
}
\end{figure}

\begin{figure}[h]
\hspace{2cm}
\psfig{file=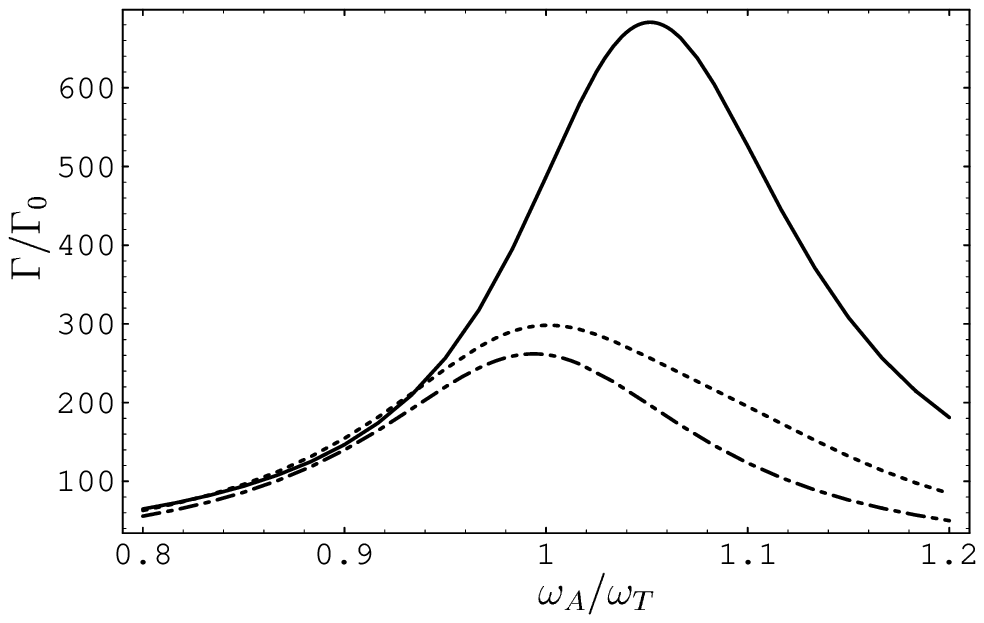,width=10cm}
\caption{\label{fig3}
The spontaneous decay rate $\Gamma$, Eq.~(\protect\ref{2.17}),
is shown as a function of the atomic transition frequency $\omega_A$
near a medium resonance for the model permittivity (\protect\ref{4.1}) 
(\mbox{$\omega_P$ $\!=$ $\!0.46\,\omega_T$}, $\gamma$ $\!=$ $\!0.2\,\omega_T$) 
and $R$ $\!=$ $\!0.02\,\lambda_A$.
The solid line corresponds to the real-cavity model, $\Gamma_{\rm GL}$
from Eq.~(\protect\ref{4.a6}), and the dotted line corresponds to
the virtual-cavity model, $\Gamma_{\rm CM}$
from Eq.~(\protect\ref{3.7}), the broken line indicating
the transverse-field assisted rate $\Gamma_{\rm CM}^\perp$
from Eq.~(\protect\ref{3.7a}).
}
\end{figure}

\begin{figure}[h]
\hspace{2cm}
\psfig{file=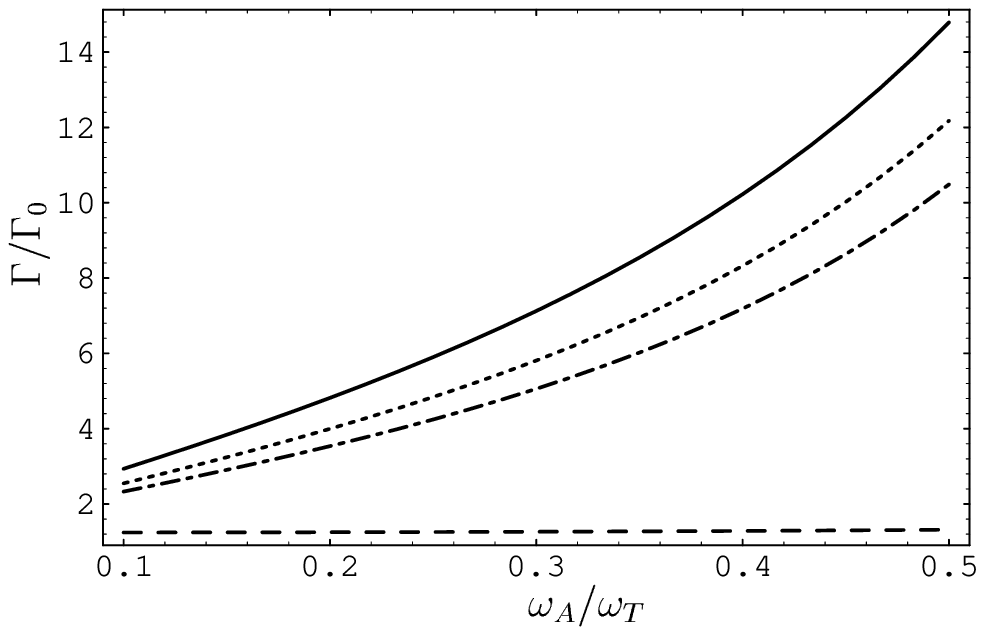,width=10cm}
\caption{\label{fig4}
The spontaneous decay rate $\Gamma$, Eq.~(\protect\ref{2.17}),
is shown as a function of the atomic transition frequency $\omega_A$
far from a medium resonance for the model permittivity (\protect\ref{4.1}) 
(\mbox{$\omega_P$ $\!=$ $\!0.46\,\omega_T$}, $\gamma$ $\!=$ $\!0.2\,\omega_T$) 
and $R$ $\!=$ $\!0.02\,\lambda_A$.
The solid line corresponds to the real-cavity model, $\Gamma_{\rm GL}$
from Eq.~(\protect\ref{4.a6}), and the dotted line corresponds to
the virtual-cavity model, $\Gamma_{\rm CM}$
from Eq.~(\protect\ref{3.7}), the broken line indicating
the transverse-field assisted rate $\Gamma_{\rm CM}^\perp$
from Eq.~(\protect\ref{3.7a}). For comparison,
the rate $\Gamma_{\rm GL}$ as obtained from 
Eq.~(\protect\ref{1.3}) together with (\protect\ref{1.5}) is
shown (dashed line).
}
\end{figure}

\begin{figure}[h]
\hspace{2cm}
\psfig{file=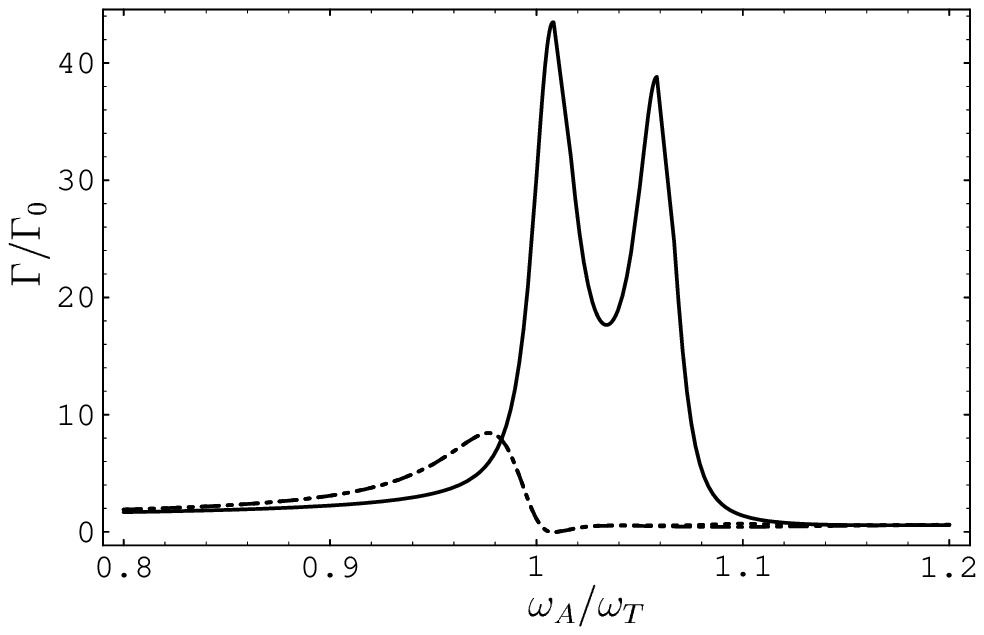,width=10cm}
\caption{\label{fig5}
The spontaneous decay rate $\Gamma$, Eq.~(\protect\ref{2.17}),
is shown as a function of the atomic transition frequency $\omega_A$
near a medium resonance for the model permittivity (\protect\ref{4.1}) 
(\mbox{$\omega_P$ $\!=$ $\!0.46\,\omega_T$}, $\gamma$ $\!=$
$\!0.05\,\omega_T$) and $R$ $\!=$ $\!0.2\,\lambda_A$.
The solid line corresponds to the real-cavity model, $\Gamma_{\rm GL}$
from Eq.~(\protect\ref{4.a6}), and the dotted line corresponds to
the virtual-cavity model, $\Gamma_{\rm CM}$
from Eq.~(\protect\ref{3.7}), the broken line indicating
the transverse-field assisted rate $\Gamma_{\rm CM}^\perp$
from Eq.~(\protect\ref{3.7a}).
}
\end{figure}

\begin{figure}[h]
\hspace{2cm}
\psfig{file=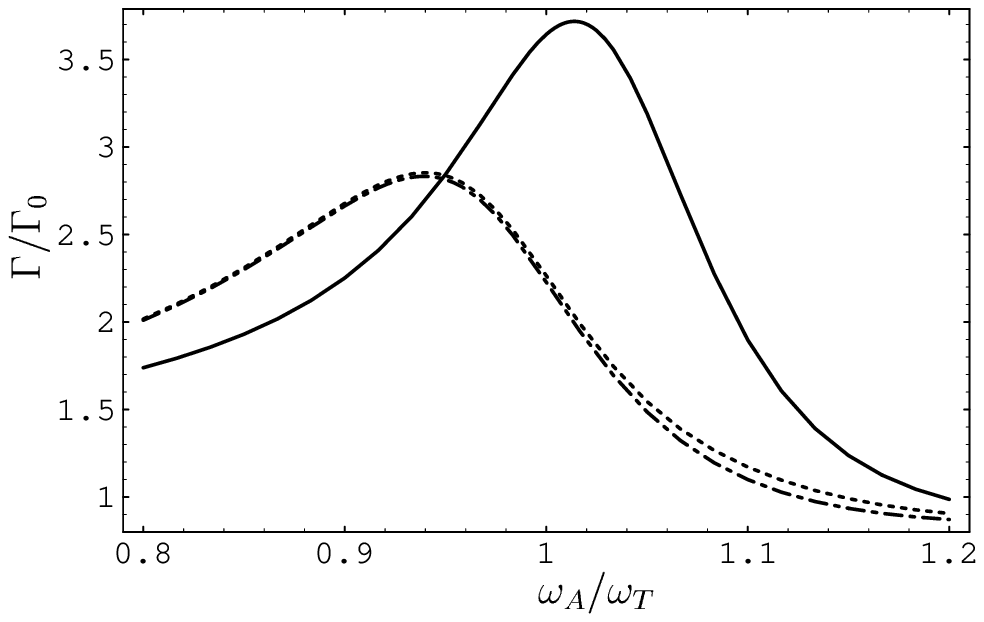,width=10cm}
\caption{\label{fig6}
The spontaneous decay rate $\Gamma$, Eq.~(\protect\ref{2.17}),
is shown as a function of the atomic transition frequency $\omega_A$
near a medium resonance for the model permittivity (\protect\ref{4.1}) 
(\mbox{$\omega_P$ $\!=$ $\!0.46\,\omega_T$}, $\gamma$ $\!=$ $\!0.2\,\omega_T$) 
and $R$ $\!=$ $\!0.2\,\lambda_A$.
The solid line corresponds to the real-cavity model, $\Gamma_{\rm GL}$
from Eq.~(\protect\ref{4.a6}), and the dotted line corresponds to
the virtual-cavity model, $\Gamma_{\rm CM}$
from Eq.~(\protect\ref{3.7}), the broken line indicating
the transverse-field assisted rate $\Gamma_{\rm CM}^\perp$
from Eq.~(\protect\ref{3.7a}).
}
\end{figure}

To illustrate the results, we have computed the (virtual cavity model)
decay rate $\Gamma_{\rm CM}$, Eq.~(\ref{3.7}) -- (\ref{3.7a}), 
and the (real cavity model) decay rate $\Gamma_{\rm GL}$, Eq.~(\ref{4.a6}),
of an atom in a spherical cavity of radius $R$ in a surrounding 
medium with the single-resonance model permittivity  
\begin{equation}
\label{4.1}
\epsilon(\omega) = 1+ \frac{\omega_P^2}
{\omega_T^2-\omega^2-i\gamma \omega_T}\,.
\end{equation}
Plots of the rates
as functions of the atomic transition frequency are 
given in Figs.~\ref{fig1} -- \ref{fig6}. The figures reveal that 
the two models can yield decay rates that are quite different 
from each other. Far from the absorption band of the
medium the difference is rather quantitative than qualitative
[Figs.~\ref{fig2} and \ref{fig4}]. 
In the absorption band and in the vicinity of the absorption band,
i.e., in the region between the medium resonance $\omega_T$ and
the longitudinal frequency $\omega_L$ $\!=$ $\!\sqrt{\omega_T^2
+ \omega_P^2}$ (in the figures, $\omega_L$ $\!=$ $1.1\,\omega_T$), 
a quantitatively and qualitatively different 
behavior of the two rates can be observed 
[Figs.~\ref{fig1}, \ref{fig3}, \ref{fig5}, and \ref{fig6}]. 
In particular, the rate obtained with the real cavity
model can substantially exceed the rate obtained with
the virtual cavity model. The differences between the
two rates are less pronounced for strong absorption;
i.e., when the value of the bandwidth parameter $\gamma$ in
Eq.~(\ref{4.1}) is sufficiently large (compare Fig.~\ref{fig1} with 
Fig.~\ref{fig3}, and Fig.~\ref{fig5} with Fig.~\ref{fig6}).
In that region the rates sensitively respond to a change of the radius
of the cavity (compare Fig.~\ref{fig1} 
with Fig.~\ref{fig5}, and Fig.~\ref{fig3} with Fig.~\ref{fig6}). 

Obviously, an excited atom in an absorbing medium undergoes both
radiative and nonradiative damping, and in dense media
nonradiative decay can be much faster than radiative one.
In particular, for small cavity radius the $\sim R^{-3}$ dipole-dipole
energy transfer terms in the two rates can strongly enhance them.
Since the radiationless decay typically
happens at the longitudinal frequency $\omega_L$,
one observes, for sufficiently small values of $\gamma$, a shift 
of the maximum of the decay rate from $\omega_T$ to $\omega_L$
with decreasing value of $R$
(compare Fig.~\ref{fig5} with Fig.~\ref{fig1}).
Even when the atomic transition frequency is relatively far from the
medium resonance, so that the imaginary part of the permittivity
becomes relatively small, the values of the two rates can  
notably differ from those obtained from Eq.~(\ref{1.3})
together with either Eq.~(\ref{1.4}) or (\ref{1.5}), because
of the $\sim R^{-3}$ near field contributions to the rates. It should 
be stressed that Eqs.~(\ref{1.3}) -- (\ref{1.5}) apply only when 
nonradiative decay can be fully excluded from consideration.
Otherwise the near-field terms can give rise to observable effects,  
as is illustrated in Figs.~\ref{fig2} and \ref{fig4}.

The rates $\Gamma_{\rm CM}$ and $\Gamma_{\rm GL}$ 
differ essentially in the way the cavity radius is introduced. 
As already mentioned, in the virtual cavity 
model the needed coincidence limit of the two spatial arguments 
of the imaginary part of the Green tensor cannot be performed,
because of the singularity of the Green tensor of the (undisturbed) medium,
and regularization is required. In the paper,  
a small fictitious distance \mbox{$|{\bf r}$ $-\!$ $\!{\bf r}'|$ 
$\!\neq$ $\!0$} between two neighboring atomic positions
inside a sphere of radius $R$ 
is kept in order to get a finite value, and the 
result is then averaged with regard to ${\bf r}$ and
${\bf r}'$ separately over the sphere. 
In contrast, in the real cavity 
model the limit ${\bf r},{\bf r}'$ $\!\to$ ${\bf r}_A$ can 
be performed exactly and a proper rate can be obtained,
$R$ being the radius of the real cavity. 
>From the above it is suggested
that the value of the parameter $R$ may be different
in the two models in order to fit each other
(note that in \mbox{Figs.~\ref{fig1} --  \ref{fig6}}
the two rates are compared for equal values of $R$). 

Another consequence of the via smoothing introduced
radius of the sphere in the virtual cavity model
is that there is a non-vanishing $\sim R^{-3}$ longitudinal-field 
contribution to the decay rate. 
Hence, the nonradiative
dipole-dipole energy transfer from the atom to the 
surrounding medium is obtained from the interaction 
of both transverse and longitudinal electromagnetic 
field components with the atom,
$\Gamma_{\rm CM}$ $\!=$ $\Gamma_{\rm CM}^\|$ $\!+$ $\Gamma_{\rm CM}^\perp$.
On the contrary, the
real cavity model leads to a decay rate that solely
results from the interaction of the atom with
the transverse field, $\Gamma_{\rm GL}$ $\!=$ $\Gamma_{\rm
GL}^\perp$. Here, the dipole-dipole energy
transfer fully corresponds to a second-order process via 
virtual photons. It is worth noting that for not too small
values of the radius of the virtual cavity (in our example,
$R$ $\!\gtsim$ $\!0.1\,\lambda_A$) the contribution
of $\Gamma_{\rm CM}^\|$ to $\Gamma_{\rm CM}$ is small,
so that it may be disregarded and hence
$\Gamma_{\rm CM}$ $\!\approx$ $\Gamma_{\rm CM}^\perp$
(see Figs.~\ref{fig5} and \ref{fig6}).

Equation (\ref{2.17}) defines the total energy
relaxation rate of the (two-level) atom, which results from both
radiative and nonradiative decay, and the question arises
of what is the spontaneous emission rate. In \cite{Barnett96}
the transverse contribution to the decay rate is 
associated with spontaneous emission, whereas the longitudinal
contribution is associated with nonradiative decay. However,
the exact result obtained with the real cavity model reveals that 
there is no longitudinal contribution to the decay rate,
and hence the transverse contribution must be
associated with both spontaneous emission and nonradiative
decay. Similarly, the decay rate obtained from the study of 
the resonant energy transfer between two guest molecules
surrounded by a perfect lattice of absorbing molecules
contains only transverse-field contributions and
describes both radiative and nonradiative relaxation
processes \cite{Juzeliunas94,Juzeliunas97}. 
In \cite{Juzeliunas97}
it is suggested that the spontaneous emission rate be identified 
with the $R$-independent (far-field) contribution to the decay
rate. Since the $\sim R^{-3}$ near field contribution
may be regarded as resulting from nonradiative decay via
dipole-dipole energy transfer, the question remains
of what is the meaning of the remaining terms. 
Moreover, from our analysis of, e.g., the 
real cavity model it is seen that $R$ must not substantially exceed
the atomic transition wavelength $\lambda_A$. Otherwise, 
the Markov approximation does not apply and the calculated decay 
rate becomes unphysical. In order to answer the question of what
is really spontaneous emission, the model should be extended
such that light detection at certain distances from the guest
atom is included.

Both in the virtual cavity model and the real cavity model
the dielectric is described in terms of a continuous polarization 
field that does not resolve the positions of the microscopic 
constituents of the medium. In reality an excited guest atom
does of course not interact with a continuous medium, but it 
``sees'' the discrete distribution of the microscopic
constituents of the medium, at least the nearest-neighbor
grouping. Hence a refined treatment of the medium
should also allow for the presence in the cavity
of nearest-neighboring medium species whose interaction 
with the guest atom is considered separately. The enlarged 
cavity can then be chosen such that the guest atom 
cannot ``resolve'' the discrete structure of the 
medium outside the cavity and the continuous description
applies \cite{Krutitsky97,Krutitsky98}.


\acknowledgements
The authors thank M. Fleischhauer for helpful comments.
S. S. likes to thank J. Audretsch, F. Burgbacher, and K. Krutitsky for
helpful discussions during his stay at the University of Konstanz.


\begin{appendix}

\section{Markov approximation}
\label{markov}

Equation (\ref{2.13}) can be formally integrated to obtain
\begin{equation}
\label{2.14}
\hat{f}_i({\bf r},\omega,t) =
\hat{f}_{{\rm free}\,i}({\bf r},\omega,t) 
+\,\frac{\omega_{21}\omega}{c^2} 
\sqrt{\frac{\epsilon_I({\bf r},\omega)}{\hbar\pi\epsilon_0}}
\, (d_{21})_k G_{ki}^\ast({\bf r}_A,{\bf r},\omega)
\int_{t'}^t d\tau\, e^{-i\omega(t-\tau)}\hat{A}_{12}(\tau),
\end{equation}
where $\hat{\bf f}_{{\rm free}}({\bf r},\omega,t)$ evolves
freely. Substituting in the vector potential in Eqs.~(\ref{2.11}) 
-- (\ref{2.12})
for $\hat{f}_i({\bf r},\omega,t)$
the expression given in Eq.~(\ref{2.14}) yields
a system of integro-differential equations for the atomic
quantities, which cannot be solved analytically in general.
Usually the Markov approximation is introduced. 
It is assumed that (after performing the $\omega$ integration) 
the time integral effectively runs over
a small correlation time interval $\tau_{\rm c}$.
As long as we require that $t$ $\!-$ $\!t'$ $\!\gg$ $\!\tau_{\rm c}$,
we may extend the lower limit of the $\tau$ integral
in Eq.~(\ref{2.14}) to minus infinity with little error.
Further we require that $\tau_{\rm c}$ be small on a time scale
on which the atomic system is changed owing to the coupling
to the electromagnetic field. In this case in the $\tau$ integral
in Eq.~(\ref{2.14}) the slowly varying atomic quantity
$\hat{A}_{12}(\tau)e^{i\omega_{21}\tau}$ can be taken at time 
$t$ and put in front of the integral,  
\begin{eqnarray}
\label{B3}
\lefteqn{
\int_{t'}^t d\tau\, e^{-i\omega(t-\tau)}\hat{A}_{12}(\tau)
\approx 
\int_{-\infty}^t d\tau\, e^{-i\omega(t-\tau)}\hat{A}_{12}(\tau)
}
\nonumber \\&&\hspace{2ex}
\approx \hat{A}_{12}(t) 
\int_{-\infty}^t \!\! d\tau \,e^{-i(\omega-\omega_{21}) (t-\tau)} 
= \hat{A}_{12}(t)\,\zeta(\omega_{21}-\omega) \nonumber \\ 
\end{eqnarray}
[$\zeta(x)$ $\!=$ $\!\pi\delta(x)$ $\!+$ $\!i{\cal P}x^{-1}$;
${\cal P}$ denotes the principal value]. Thus, the future of
the system is now determined by the present time only. 
We substitute in Eq.~(\ref{2.14}) for the time integral
the expression given in Eq.~(\ref{B3}), calculate the vector
potential, Eqs.~(\ref{2.8}) and (\ref{2.9}). 
With the help of the relation (see, e.g., \cite{Scheel98a}) 
\begin{equation}
\label{B4}
\int d^3{\bf s}\;
\frac{\omega^2}{c^2}\, \epsilon_I({\bf s},\omega)
G_{km}({\bf r},{\bf s},\omega) G^\ast_{lm}({\bf r'},{\bf s},\omega) 
= {\rm Im}G_{kl}({\bf r},{\bf r'},\omega)
\end{equation}
we find after some calculation
\begin{equation}
\label{B5}
\hat{A}_i^{(+)}({\bf r}_A,t) 
= \hat{A}_{{\rm free}\,i}^{(+)}({\bf r}_A,t)
+ \,\frac{\omega_{21}}{\pi\epsilon_0 c^2} (d_{21})_k 
\int_0^\infty \!\!d\omega\,\zeta(\omega_{21}\!-\!\omega) 
{\rm Im}\,G_{ik}({\bf r}_A,{\bf r}_A,\omega) \,\hat{A}_{12}(t) .
\end{equation}
In order to obtain Eqs.~(\ref{2.15}) -- (\ref{2.16}), we
eventually substitute in Eqs.~(\ref{2.11}) -- (\ref{2.12})
for the positive and negative frequency parts of the 
vector potential the expressions according to Eq.~(\ref{B5}).
It can be easily seen that the real part of the $\zeta$ function
(i.e., the $\delta$ function) in Eq.~(\ref{B5}) leads to
$\Gamma$ given in Eq.~(\ref{2.17}).
The principal-value integral in Eq.~(\ref{B5})
which arises from the imaginary part of the $\zeta$ function 
contributes to the Lamb shift and reads
\begin{eqnarray}
\label{B6}
\delta\omega=\frac{2\omega_{21}^2 (d_{21})_k (d_{21})_{k'}}{\hbar
\epsilon_0 c^2 \pi} \!\!\int_0^\infty\!\! d\omega \,
\frac{{\rm Im}\,G_{kk'}({\bf r}_A,{\bf r}_A,\omega)}{\omega -\omega_{21}}\,, 
\end{eqnarray}
which can be rewritten as
\begin{equation}
\label{B7}
\delta\omega=\frac{2\omega_{21}^2 (d_{21})_k
(d_{21})_{k'}}{\hbar\epsilon_0 c^2} 
\bigg[{\rm Re}\,G_{kk'}({\bf r}_A,{\bf r}_A,\omega_{21})
-\,\frac{1}{\pi} \int_0^\infty d\omega \, 
\frac{{\rm Im}\,G_{ik}({\bf r}_A,{\bf r}_A,\omega)}{\omega 
+\omega_{21}} \bigg].
\end{equation}
Equation (\ref{B7}) holds because of the Kramers-Kronig relation (or
Titchmarsh's theorem) for the Green function. Note that
the real part of the vacuum Green function is infinite 
for ${\bf r}$ $\!=$ $\!{\bf r}'$ $\!=$ $\!{\bf r}_A$ and regularization 
is required. The resulting
vacuum Lamb shift may be thought of as being included in the atomic 
transition frequency, so that $\delta\omega$ in Eq.~(\ref{2.16})
may be regarded as being solely due to the surrounding dielectric.


\section{Green tensor for a homogeneous dielectric}
\label{greenbulk}
Following \cite{Barnett96,Abrikosov}, the
Green tensor for bulk material can be given by
\begin{equation}
\label{C1}
G_{kk'}({\bf r},{\bf r}',\omega)
= G_{kk'}^\|({\bf r},{\bf r}',\omega)
+ G_{kk'}^\perp({\bf r},{\bf r}',\omega),
\end{equation}
where ({\boldmath$\rho$} $\!=$ ${\bf r}$ $\!-$ $\!{\bf r}'$)         
\begin{equation}
\label{C2}
G_{kk'}^\|({\bf r},{\bf r}',\omega)
= - \frac{c^2}{4\pi\omega^2\epsilon(\omega)} \left[
\frac{4\pi}{3}\delta(\mbox{\boldmath $\rho$})\,\delta_{kk'}
+ \left(\delta_{kk'}-\frac{3\rho_k\rho_{k'}}{\rho^2}\right)\frac{1}{\rho^3}
\right]
\end{equation}
and
\begin{eqnarray}
\label{C3}
G_{kk'}^\perp({\bf r},{\bf r}',\omega)
&=& \frac{c^2}{4\pi\omega^2\epsilon(\omega)}
\bigg\{ 
\left(\delta_{kk'}-\frac{3\rho_k\rho_{k'}}{\rho^2}\right)\frac{1}{\rho^3}
+\,k^3\bigg[
\left(\frac{1}{k\rho}+\frac{i}{(k\rho)^2}-\frac{1}{(k\rho)^3} \right) 
\delta_{kk'} 
\nonumber\\&&\hspace{2ex} 
- \left(\frac{1}{k\rho}+\frac{3i}{(k\rho)^2}-\frac{3}{(k\rho)^3} \right)
\frac{\rho_k\rho_{k'}}{\rho^2}
\bigg] e^{ik\rho}
\bigg\},
\end{eqnarray}
are related to the longitudinal and transverse electric fields.
In Eq.~(\ref{C3}), the complex wave number
\begin{equation}
\label{C4}
k = \sqrt{\epsilon(\omega)}\,\frac{\omega}{c}
= \left[\eta(\omega)+i\kappa(\omega)\right]\frac{\omega}{c}
\end{equation}
has been introduced.
In particular for small values of $|k\rho|$,
$|k\rho|$ $\!\ll$ $\!1$, the exponential
$e^{ik\rho}$ in Eq.~(\ref{C3}) can be expanded to obtain
\begin{equation} 
\label{C5}
G^\perp_{kk'}({\bf r},{\bf r}',\omega)
= \frac{1}{4\pi} 
\left\{ 
\frac{\rho_k \rho_{k'}}{2\rho^3} +\frac{\delta_{kk'}}{2\rho}
+ \, \frac{2i\omega}{3c} \left[ \eta(\omega) +i\kappa(\omega) \right]
\delta_{kk'} \right\} +{\cal O}(\rho).
\end{equation}


\section{Green tensor for an empty sphere surrounded by 
a homogeneous dielectric}
\label{green}

Following \cite{Li94},
the Green tensor of a system that consists of 
an empty sphere surrounded by a homogeneous dielectric 
can be given in terms of spherical Bessel functions and spherical
harmonics. When ${\bf r}$ and ${\bf r}'$ lie
in the sphere (with the center of the sphere being the origin of 
the coordinate system), then the associated Green tensor 
{\boldmath $G$}$({\bf r},{\bf r}',\omega)$
is given by
\begin{equation} 
\label{A1}
\mbox{\boldmath $G$}({\bf r},{\bf r'},\omega) 
= \mbox{\boldmath $G$}^{\rm V}({\bf r},{\bf r'},\omega) 
+ \tilde{\mbox{\boldmath $G$}}({\bf r},{\bf r'},\omega),
\end{equation}
where \mbox{\boldmath $G$}$^{\rm V}({\bf r},{\bf r'},\omega)$
is the vacuum Green tensor, and
\begin{eqnarray}
\label{A1a}
\lefteqn{
\tilde{\mbox{\boldmath $G$}}({\bf r},{\bf r'},\omega)
= \frac{i\omega}{4\pi c} \sum\limits_{e,o} \sum\limits_{n=1}^\infty
\sum\limits_{m=0}^n 
\bigg\{ 
\frac{2n\!+\!1}{n(n\!+\!1)} \frac{(n\!-\!m)!}{(n\!+\!m)!} 
} 
\nonumber \\ &&\hspace{2ex}\times \,
(2\!-\!\delta_{0m})\left[ C^M_n(\omega) {\bf M}_{{e \atop o}nm} 
\left({\bf r},\frac{\omega}{c}\right) 
{\bf M}_{{e \atop o}nm}\left({\bf r'},\frac{\omega}{c}\right) 
+\, C^N_n(\omega) {\bf N}_{{e \atop o}nm}\left({\bf r},\frac{\omega}{c}\right) 
{\bf N}_{{e \atop o}nm}\left({\bf r'},\frac{\omega}{c}\right) \right]
\bigg\}.  \nonumber \\
\end{eqnarray}
Here ${\bf M}_{{e \atop o}nm}({\bf r},k)$ and ${\bf N}_{{e \atop
o}nm}({\bf r},k)$ are the ({\em e}ven and {\em o}dd)
vector Debye potentials, and the quantities $C^{M,N}_n(\omega)$
are the generalized reflection coefficients.
Introducing the abbreviating notations 
\begin{eqnarray}
J_{ni} &=& j_n(k_i R), 
\\ 
H_{ni} &=& h_n^{(1)}(k_i R), 
\\
J'_{ni} &=& \frac{1}{\rho} \left. \frac{d[\rho j_n(\rho)]}{d\rho}
\right|_{\rho=k_i R}, 
\\
H'_{ni} &=& \frac{1}{\rho} \left. \frac{d[\rho
h_n^{(1)}(\rho)]}{d\rho}
\right|_{\rho=k_i R}
\end{eqnarray}
($k_1$ $\!=$ $\!\sqrt{\epsilon(\omega)}\,\omega/c$, $k_2$ $\!=$
$\!\omega/c$), 
the reflection coefficients can be given by
\begin{equation} 
\label{A2}
C^{M,N}_n(\omega) = \frac{T^{H,V}_{F,n}(\omega)
R^{H,V}_{P,n}(\omega)}{T^{H,V}_{P,n}(\omega)} \,,
\end{equation}
where  
\begin{eqnarray}
R^H_{P,n}(\omega) &=& \frac{k_2 H'_{n2} H_{n1} - k_1 H'_{n1} H_{n2}}{k_2
J_{n1} H'_{n2} -k_1 J'_{n1} H_{n2}}\,, \\
R^V_{P,n}(\omega) &=& \frac{k_2 H_{n2} H'_{n1} - k_1 H_{n1} H'_{n2}}{k_2
J'_{n1} H_{n2} -k_1 J_{n1} H'_{n2}}\,, \\
T^H_{P,n}(\omega) &=& \frac{k_2 [ J_{n2} H'_{n2} - J'_{n2} H_{n2}]}{k_2
J_{n1} H'_{n2} - k_1 J'_{n1} H_{n2}}\,, \\
T^H_{F,n(\omega)} &=& \frac{k_2 [ J'_{n2} H_{n2} - J_{n2} H'_{n2}]}{k_2
J'_{n2} H_{n1} - k_1 J_{n2} H'_{n1}}\,, \\
T^V_{P,n}(\omega) &=& \frac{k_2 [ J'_{n2} H_{n2} - J_{n2} H'_{n2}]}{k_2
J'_{n1} H_{n2} - k_1 J_{n1} H'_{n2}}\,, \\
T^H_{F,n}(\omega) &=& \frac{k_2 [ J_{n2} H'_{n2} - J'_{n2} H_{n2}]}{k_2
J_{n2} H'_{n1} - k_1 J'_{n2} H_{n1}}\,.
\end{eqnarray}
The vector Debye potentials are defined by
\begin{eqnarray} \label{AA1}
{\bf M}_{{e \atop o}nm}({\bf r},k) &=& {\bf \nabla} \times \left[
\psi_{{e \atop o}nm}({\bf r},k) {\bf r}
\right] ,\\ \label{AA2}
{\bf N}_{{e \atop o}nm}({\bf r},k) &=& \frac{1}{k}{\bf \nabla} \times {\bf
\nabla} \times \left[ \psi_{{e \atop o}nm}({\bf r},k) {\bf r} \right] 
\end{eqnarray}
with
\begin{equation}
\psi_{{e \atop o}nm}({\bf r},k) = j_n(kr) P_n^m(\cos\theta) {\cos
\choose \sin} m\phi ,
\end{equation}
and can be given by
\begin{equation} 
\label{A3}
{\bf M}_{{e \atop o}nm}({\bf r},k) = \frac{im}{\sin\theta} \, j_n(kr)
P_n^m(\cos\theta) {\cos \choose \sin} m\phi \,{\bf e}_{\theta}
-\,j_n(kr)\, \frac{dP_n^m(\cos\theta)}{d\theta} 
\,{\cos \choose \sin} m\phi \,{\bf e}_{\phi}, 
\end{equation}
\begin{eqnarray}
\label{A4}
\lefteqn{
{\bf N}_{{e \atop o}nm}({\bf r},k) = \frac{n(n+1)}{kr}\, j_n(kr) 
P_n^m(\cos\theta) {\cos \choose \sin} m\phi \, {\bf e}_r 
}
\nonumber \\ && \hspace{2ex}
+\,\frac{1}{kr} \frac{d[rj_n(kr)]}{dr} \left[
\frac{dP_n^m(\cos\theta)}{d\theta} {\cos \choose \sin} m\phi \, {\bf
e}_{\theta} 
\mp \,\frac{im}{\sin\theta}\, P_n^m(\cos\theta) {\sin \choose \cos}
m\phi \, {\bf
e}_{\phi} \right] ,
\end{eqnarray}
$j_n(kr)$ is the spherical Bessel function
of the first kind and $P_n^m(\cos\theta)$ is the associated
Legendre polynomial.
Note that from Eqs.~(\ref{AA1}) and (\ref{AA2}) it follows that
$\mbox{\boldmath $G$}({\bf r},{\bf r}',\omega)$ is a (two-sided)
transverse tensor function.

Since for $kr$ $\!\to$ $\!0$ we have
\begin{equation}
j_n(kr) \stackrel{kr\to 0}{\longrightarrow} \frac{(kr)^n}{(2n+1)!!}
\left( 1-\frac{1}{2(2n+3)} +\ldots \right),
\end{equation}
from inspection of Eqs.~(\ref{A3}) and (\ref{A4}) we find that 
\begin{eqnarray}
{\bf M}_{{e \atop o}nm}({\bf r},k) &\stackrel{kr\to
0}{\longrightarrow}& (kr)^n, \\ 
{\bf N}_{{e \atop o}nm}({\bf r},k) &\stackrel{kr\to
0}{\longrightarrow}& (kr)^{n-1}. 
\end{eqnarray}
Hence, at the center of the sphere only the TM-wave
vector Debye potentials ${\bf N}_{{e \atop o}10}({\bf r},k)$
and ${\bf N}_{{e \atop o}11}({\bf r},k)$ contribute to 
$\tilde{\mbox{\boldmath $G$}}({\bf r},{\bf r'},\omega)$
in Eq.~(\ref{A1a}),
\begin{equation} \label{A5}
\left.\tilde{G}_{kk'}({\bf r},{\bf r}',\omega)\right|_{{\bf r}={\bf r}'=0} 
= \frac{i\omega}{6\pi c} C^N_1(\omega) \delta_{kk'} ,
\end{equation}
where $[n$ $\!\equiv$ $\!\sqrt{\epsilon(\omega)}]$
\begin{equation} 
\label{A6}
C^N_1(\omega) = 
\frac{\left[ i+z(n+1) - iz^2n - z^3 n^2/(n+1) \right] e^{iz}}
{\sin z - z(\cos z + in \sin z) + iz^2 n \cos z 
- z^3 (\cos z -in \sin z) n^2/(n^2-1) }
\end{equation}
with 
\begin{equation}
z = \frac{R\omega}{c} \,.
\end{equation}

\end{appendix}

\begin{references}
\bibitem{note}
Note that in \cite{Scheel98b} the average is taken with respect
to the distance ${\bf r}$ $\!-$ $\!{\bf r}_A$, which leads
to somewhat different regularization factors.
\end{references}
\end{document}